\documentclass[aps,prl,twocolumn,floatfix,superscriptaddress]{revtex4-1}
\usepackage[encapsulated]{CJK}
\pdfoutput=1
\usepackage{graphicx}
\usepackage{amsmath,amsfonts,amssymb}
\usepackage[colorlinks=true,linkcolor=blue,citecolor=blue]{hyperref}
\usepackage{siunitx}
\usepackage{color,soul}

\begin{document}
\begin{CJK*}{UTF8}{}
\title{Global strain-induced scalar potential in graphene devices}
\author{Lujun Wang}
\email{lujun.wang@unibas.ch}
\affiliation{Department of Physics, University of Basel, Klingelbergstrasse 82, CH-4056 Basel, Switzerland}
\affiliation{Swiss Nanoscience Institute, University of Basel, Klingelbergstrasse 82, CH-4056 Basel, Switzerland}

\author{Andreas Baumgartner}
\affiliation{Department of Physics, University of Basel, Klingelbergstrasse 82, CH-4056 Basel, Switzerland}
\affiliation{Swiss Nanoscience Institute, University of Basel, Klingelbergstrasse 82, CH-4056 Basel, Switzerland}

\author{P\'eter Makk}
\affiliation{Department of Physics, University of Basel, Klingelbergstrasse 82, CH-4056 Basel, Switzerland}
\affiliation{Department of Physics, Budapest University of Technology and Economics and Nanoelectronics Momentum Research Group of the Hungarian Academy of Sciences, Budafoki ut 8, 1111 Budapest, Hungary}

\author{Simon Zihlmann}
\affiliation{Department of Physics, University of Basel, Klingelbergstrasse 82, CH-4056 Basel, Switzerland}

\author{Blesson S. Varghese}
\affiliation{Department of Physics, University of Basel, Klingelbergstrasse 82, CH-4056 Basel, Switzerland}

\author{David I. Indolese}
\affiliation{Department of Physics, University of Basel, Klingelbergstrasse 82, CH-4056 Basel, Switzerland}

\author{Kenji Watanabe}
\affiliation{Research Center for Functional Materials, National Institute for Material Science, 1-1 Namiki, Tsukuba, 305-0044, Japan}

\author{Takashi Taniguchi}
\affiliation{International Center for Materials Nanoarchitectonics, National Institute for Materials Science, 1-1 Namiki, Tsukuba 305-0044, Japan}

\author{Christian Sch\"onenberger}
\affiliation{Department of Physics, University of Basel, Klingelbergstrasse 82, CH-4056 Basel, Switzerland}
\affiliation{Swiss Nanoscience Institute, University of Basel, Klingelbergstrasse 82, CH-4056 Basel, Switzerland}

\begin{abstract}
By mechanically distorting a crystal lattice it is possible to engineer the electronic and optical properties of a material. In graphene, one of the major effects of such a distortion is an energy shift of the Dirac point, often described as a scalar potential. We demonstrate how such a scalar potential can be generated systematically over an entire electronic device and how the resulting changes in the graphene work function can be detected in transport experiments. Combined with Raman spectroscopy, we obtain a characteristic scalar potential consistent with recent theoretical estimates. This direct evidence for a scalar potential on a macroscopic scale due to deterministically generated strain in graphene paves the way for engineering the optical and electronic properties of graphene and similar materials by using external strain.

\end{abstract}

\maketitle
\end{CJK*}

Graphene is a model system on which a large variety of new and prominent physical phenomena have been discovered~\cite{CastroNeto2009, DasSarma2011, Goerbig2011, Yankowitz2019}. A particularly promising topic is the control of its electronic properties by external strain, which has been extensively studied theoretically. The predicted strain effects in the low-energy band structure of graphene can be summarized as changes in the magnitude and isotropy of the Fermi velocity and thus in the density of states (DoS)~\cite{Pereira2009a, Choi2010, Juan2012, Grassano2020}, shifts in the energy of the Dirac point, which is typically incorporated as a scalar potential~\cite{Guinea2010, Choi2010, Grassano2020}, and changes in the position of the Dirac cone in the two-dimensional Brillouin zone, often described by a pseudo-vector potential acting on the valley degree of freedom~\cite{Fogler2008, Guinea2009, Guinea2010, Low2010, Uchoa2013, Zhu2015}. Previous experiments explored some of these strain effects on a local scale using scanning tunneling microscopy~\cite{Levy2010, Klimov2012, Yan2012, Guo2012, Lu2012, Jiang2017, Liu2018, Jia2019, Li2020}, Kelvin probe force microscopy~\cite{He2015, Volodin2017}, or angle-resolved photoemission spectroscopy~\cite{Nigge2019}. However, studying strain effects in transport measurements and on a global scale is still challenging due to the lack of {\it in situ} strain tunability~\cite{Shioya2015, Wu2018, Liu2018} or ambiguities resulting from simultaneous changes in the gate capacitance ~\cite{Huang2011, Guan2017, Wang2019a}.

Here, we demonstrate the formation of a scalar potential generated by systematically tuning the strain in a micrometer sized graphene electronic device and investigate its effects on two fundamental electron transport phenomena, quasi-ballistic transport and the quantum Hall effect (QHE). We find that all investigated transport characteristics are shifted systematically in gate voltage, qualitatively and quantitatively consistent with the expectations for the scalar potential generated by the applied strain, where the strain values are confirmed by separate Raman spectroscopy experiments.

\begin{figure}[htb]
	\centering
	\includegraphics[]{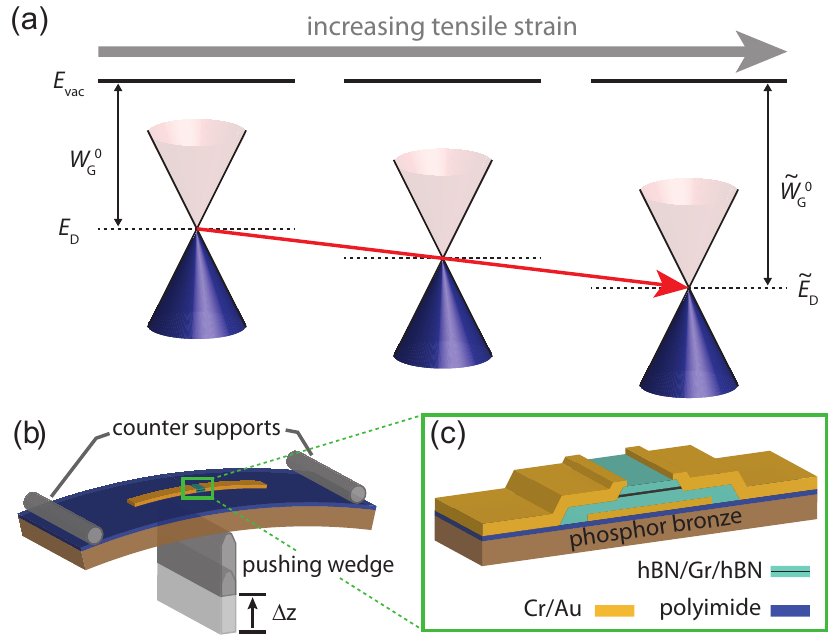}
	\caption{\textbf{(a)} Illustration of the strain induced shift of the Dirac point. The vacuum level is labeled with $E_{\rm vac}$, and the Dirac point and the work function of unstrained (strained) graphene with $E_{\rm D}$ ($\widetilde{E}_{\rm D}$) and $W_{\rm G}^{0}$ ($\widetilde{W}_{\rm G}^{0}$), respectively. \textbf{(b)} Schematics of the three-point bending setup and \textbf{(c)} the encapsulated graphene device. The displacement $\mathrm{\Delta}z$ of the pushing wedge controls the bending of the substrate and thus the induced strain in the graphene.}
	\label{fig:fig1}
\end{figure}

The work function (WF) of a material, i.e. the energy required to remove an electron from the material, is defined as the difference between the vacuum level $E_{\rm vac}$ and the Fermi level $E_{\rm F}$ of the material~\cite{Cahen2003}. For undoped graphene, $E_{\rm F}$ coincides with the Dirac point energy $E_{\rm D}$~\cite{CastroNeto2009}, therefore the WF of undoped graphene is $W_{\rm G}^{0} = E_{\rm vac} - E_{\rm D}$. A strain-induced scalar potential shifts $E_{\rm D}$, which therefore leads to a change in $W_{\rm G}^{0}$. With increasing tensile strain, the scalar potential shifts $E_{\rm D}$ to lower values, resulting in an increase in $W_{\rm G}^{0}$~\cite{Choi2010, Grassano2020}, as illustrated in Fig.~\ref{fig:fig1}(a).
Quantitatively, strain shifts $E_{\rm D}$ to $\widetilde{E}_{\rm D} = E_{\rm D} + S$, where $S$ is the scalar potential, and can be written as~\cite{Choi2010, Guinea2010, Grassano2020}:
\begin{equation}\label{eq:Scalar_potential}
S(x,y) = -s_0 \cdot (\varepsilon_\text{xx}+\varepsilon_\text{yy}),
\end{equation}
with $\varepsilon_\text{xx}$ and $\varepsilon_\text{yy}$ the diagonal components of the strain tensor, and $s_0$ a constant defined for small strain values. The value of $s_0$ is not well established and theoretical values are reported in the range between \SI{2.5}{\eV} and \SI{4.1}{\eV}~\cite{Choi2010, Guinea2010, Grassano2020}.

How we generate strain in our experiments in an on-chip fully encapsulated graphene device is illustrated in Fig.~\ref{fig:fig1}(b): in a three-point bending setup a $\SI{24}{mm} \times \SI{9.5}{mm} \times \SI{0.3}{mm}$ flexible substrate with the devices fabricated in the center is bent by pushing a central wedge against two fixed counter supports by a displacement of $\mathrm{\Delta}z$ ~\cite{Wang2019a}. The schematics of the device configuration is shown in Fig.~\ref{fig:fig1}(c). The edge contacts to graphene act as clamps for the strain generation and at the same time as electrical contacts for transport experiments~\cite{Wang2019a}. A metallic global bottom gate is used to tune the charge carrier density in the device. The on-chip hBN encapsulation ensures that the geometrical capacitance between the gate and the graphene is not changed in the straining process. Here, we investigate strain effects on devices with a rectangular geometry, which results in an essentially homogeneous uniaxial strain field. Details of the device fabrication and the strain field pattern are discussed in~\cite{Wang2019a}.


\begin{figure}[htb]
	\centering
	\includegraphics[]{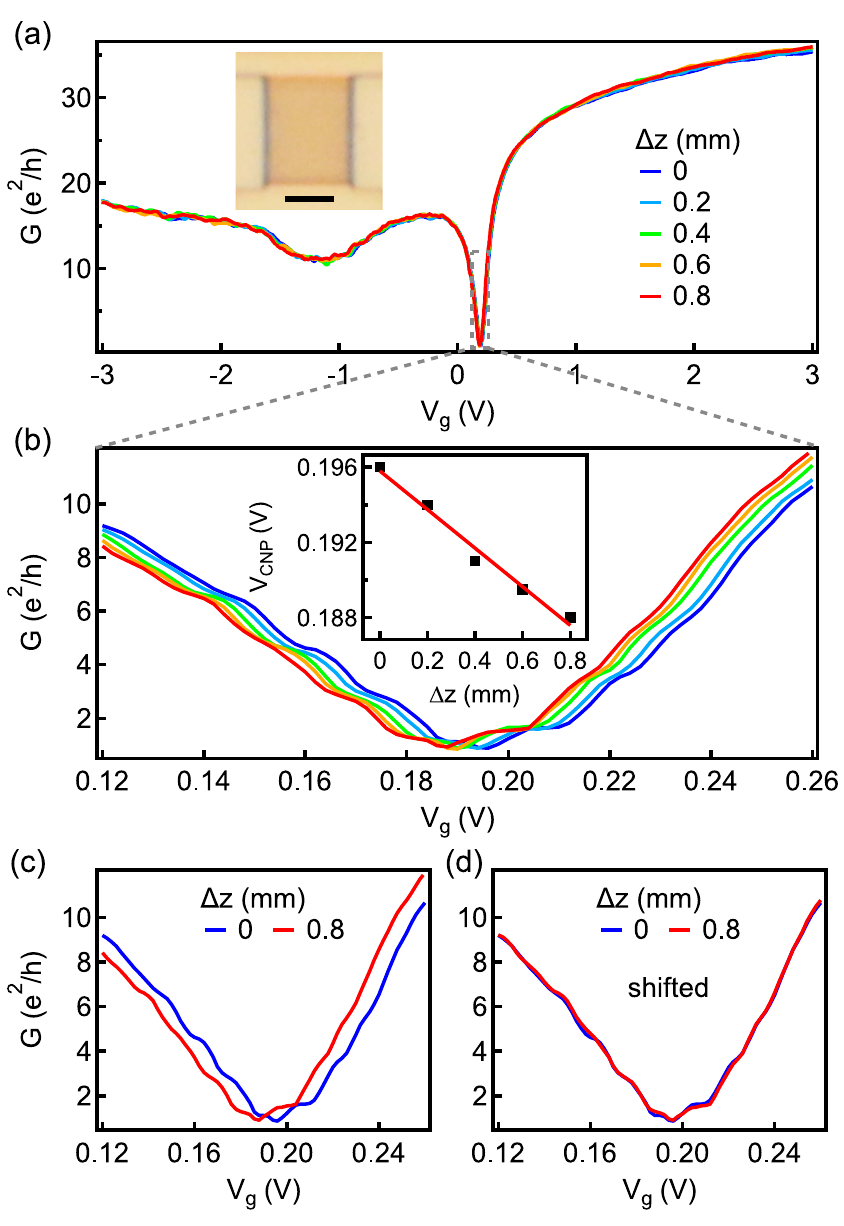}
	\caption{\textbf{(a)} Two-terminal differential conductance $G$ as a function of the gate voltage $V_\text{g}$ for different $\mathrm{\Delta}z$. Inset: optical image of the measured device, scale bar: \SI{2}{\micro\meter}. \textbf{(b)} Zoom-in on the CNP. Inset: position of the CNP ($V_\text{CNP}$) as a function of $\mathrm{\Delta}z$. $V_\text{CNP}$ is extracted as the gate voltage of minimum conductance. Red line is a linear fit with a slope of about \SI{-10}{mV/mm}. \textbf{(c)} Zoom-in on the CNP for $\mathrm{\Delta}z=0$ and $\mathrm{\Delta}z=\SI{0.8}{mm}$. \textbf{(d)} Same data as in \textbf{(c)} with the $\mathrm{\Delta}z=\SI{0.8}{mm}$ curve (red) shifted to the right by \SI{+8}{mV} in $V_\text{g}$.}
	\label{fig:fig2}
\end{figure}

In our devices, the grounded graphene sheet and the metallic gate essentially form a plate capacitor. The detailed diagram of energy level alignment and its modification by strain are given in the Supplemental Material. The strain-induced scalar potential shifts the Dirac point, resulting in a systematic change in the charge carrier density of the device at a given gate voltage, which we detect in transport experiments.

To investigate the strain effect, we perform transport experiments at liquid helium temperature ($T\approx\SI{4.2}{\kelvin}$) using standard low-frequency lock-in techniques. The two-terminal differential conductance $G=dI/dV$ of a square device is measured as a function of $V_\text{g}$ for different bendings $\mathrm{\Delta}z$ of the substrate. An overview measurement is plotted in Fig.~\ref{fig:fig2}(a), on the scale of which no significant strain effects can be observed. The charge neutrality point (CNP) occurs at a positive gate voltage.
From a linear fit near the CNP we find a field effect charge carrier mobility of $\SI{\sim 130000}{\square\centi\meter\per\volt\per\second}$, independent of $\mathrm{\Delta}z$, suggesting a high device quality and that random strain fluctuations are probably not dominating scattering processes here~\cite{Wang2020}. The additional conductance minimum at $V_\text{g}\approx\SI{-1.2}{\volt}$ may originate from a large contact doping due to the overlap of the electrodes with the graphene region near the edge contacts~\cite{Du2018}, or from a super-superlattice effect in encapsulated graphene when both the top and the bottom hBN layers are aligned to the graphene lattice~\cite{Wang2019}.


The zoom-in to the CNP plotted in Fig.~\ref{fig:fig2}(b) shows very regular oscillations in conductance, which we tentatively attribute to Fabry-P\'erot resonances in the regions near the electrical contacts with a different doping compared to the graphene bulk~\cite{Young2009, Rickhaus2013, Grushina2013, Handschin2017} (see Supplemental Material for a detailed discussion).
%
%
%
With increasing $\mathrm{\Delta}z$ and therefore increasing tensile strain, these conductance oscillations are shifted systematically to lower gate voltages. This effect is fully reversible with deceasing $\mathrm{\Delta}z$, which is demonstrated in the Supplemental Material. The strain-induced shift is best seen by following the CNP: in the inset of Fig.~\ref{fig:fig2}(b) we plot the gate voltage of minimum conductance, $V_\text{CNP}$ as a function of $\mathrm{\Delta}z$, which shows a linear decrease with increasing $\mathrm{\Delta}z$, consistent with the picture described in the Supplemental Material. To demonstrate that the complete conductance curves are shifted with strain, we plot in Fig.~\ref{fig:fig2}(c) the two curves with the lowest ($\mathrm{\Delta}z=0$) and the highest ($\mathrm{\Delta}z=\SI{0.8}{\milli\meter}$) strain values, and in Fig.~\ref{fig:fig2}(d) the same data, but with the $\mathrm{\Delta}z=\SI{0.8}{\milli\meter}$ curve (red) shifted by \SI{+8}{mV} in $V_\text{g}$. We find that all conductance curves merge to the same curve as at $\mathrm{\Delta}z=0$ (blue) when shifted by a constant gate voltage offset. This shift we attribute to a strain-induced scalar potential in the graphene sheet. We note that this effect is very different from bending-induced changes in the gate capacitance found in suspended samples, where the $V_\text{g}$ axis is rescaled by a constant factor~\cite{Wang2019a}.

\begin{figure}[htb]
	\centering
	\includegraphics[]{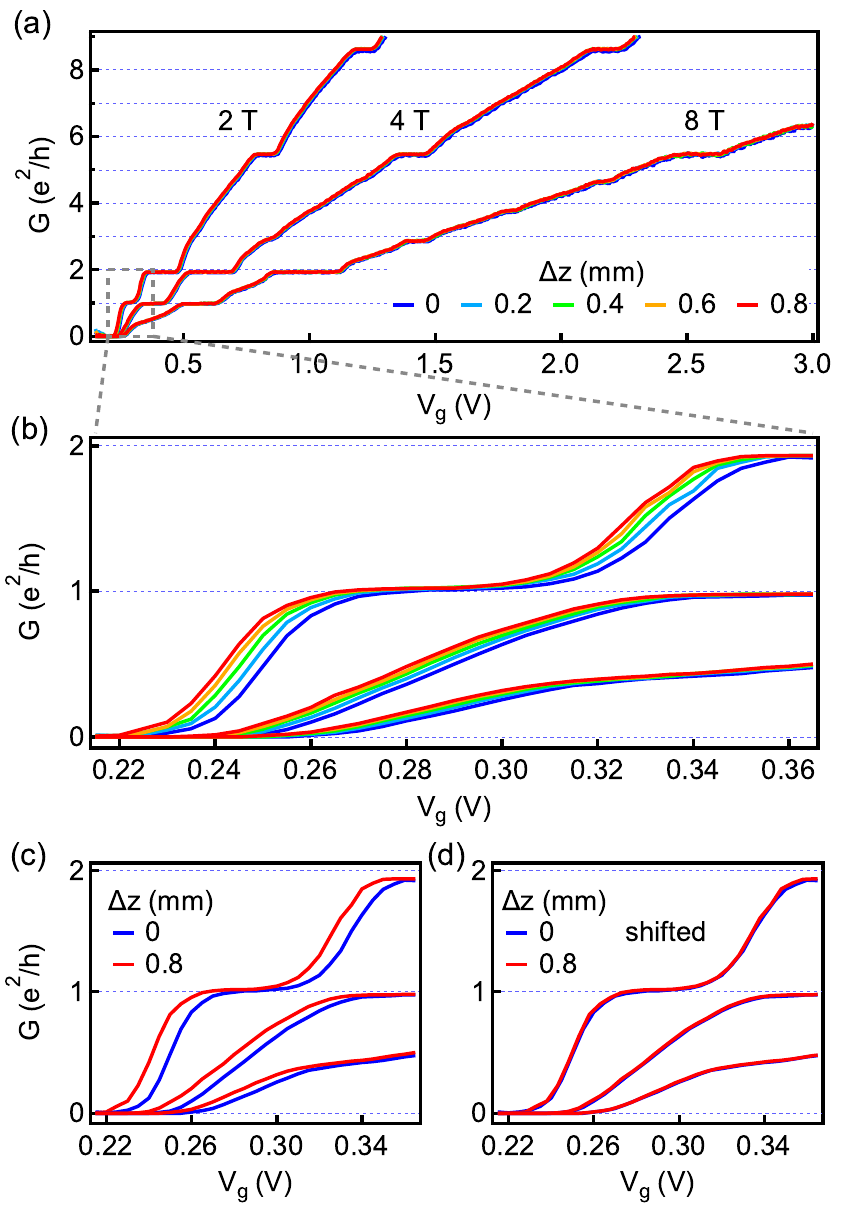}
	\caption{\textbf{(a)} Two-terminal differential conductance as a function of gate voltage at three different magnetic fields for different $\mathrm{\Delta}z$. \textbf{(b)} Zoom-in to a small region in \textbf{(a)}, showing the shift of the curves in $V_\text{g}$ with increasing $\mathrm{\Delta}z$. \textbf{(c)} Same as in \textbf{(b)} for $\mathrm{\Delta}z=0$ and $\mathrm{\Delta}z=\SI{0.8}{mm}$. \textbf{(d)} Same as in \textbf{(c)} with the $\mathrm{\Delta}z=\SI{0.8}{mm}$ curve (red) shifted to the right by \SI{8}{mV} in $V_\text{g}$.}
	\label{fig:fig3}
\end{figure}

To demonstrate that this is a general effect, independent of the device or the physical origin of the transport characteristics, we have investigated more than 5 devices, all showing similar effects (another example is provided in the Supplemental Material). Here, we now focus on the impact of homogeneous uniaxial strain on the QHE in the same device, and perform a similar analysis as for the zero field measurements. Figure~\ref{fig:fig3}(a) shows the two-terminal differential conductance as a function of the gate voltage for three different quantizing magnetic fields, $B$, and for different $\mathrm{\Delta}z$ values. Typical quantum Hall plateaus of graphene can be observed on the electron side, with small deviations of the plateau conductances from the quantized values 2, 6, 10$\,e^2/h$ due to the contact resistance. The plateaus at the filling factors $\nu=0$ and $\nu=1$ are well developed alreday at $B=\SI{2}{\tesla}$, and more broken symmetry states and fractional quantum Hall states can be observed at $B=\SI{8}{\tesla}$~\cite{Bolotin2009, Du2009, Dean2011}, again highlighting the very good device quality.
In contrast, the plateaus on the hole side are not well developed (see Supplemental Material) presumably due to a p-n junction forming near the contacts~\cite{Oezyilmaz2007, Amet2014}. Comparing the measurements for different $\mathrm{\Delta}z$ on this scale shows no clear strain effects. However, in the data near the CNP shown in Fig.~\ref{fig:fig3}(b), we again find a systematic shift in $V_\text{g}$ with increasing $\mathrm{\Delta}z$. The clear offset between the $\mathrm{\Delta}z=\SI{0.8}{mm}$ curve (red) and the $\mathrm{\Delta}z=0$ curve (blue) is shown in Fig.~\ref{fig:fig3}(c). Shifting the red curve by \SI{+8}{mV}, as shown in Fig.~\ref{fig:fig3}(d), the two curves are virtually identical, in the same manner and with the same shift as discussed for Fig.~\ref{fig:fig2} with the device at zero magnetic field. Since the QHE is quite a different transport regime than quasi-ballistic transport, the observed effect is very general and we attribute it to a strain-induced scalar potential.

\begin{figure}[htb]
	\centering
	\includegraphics[]{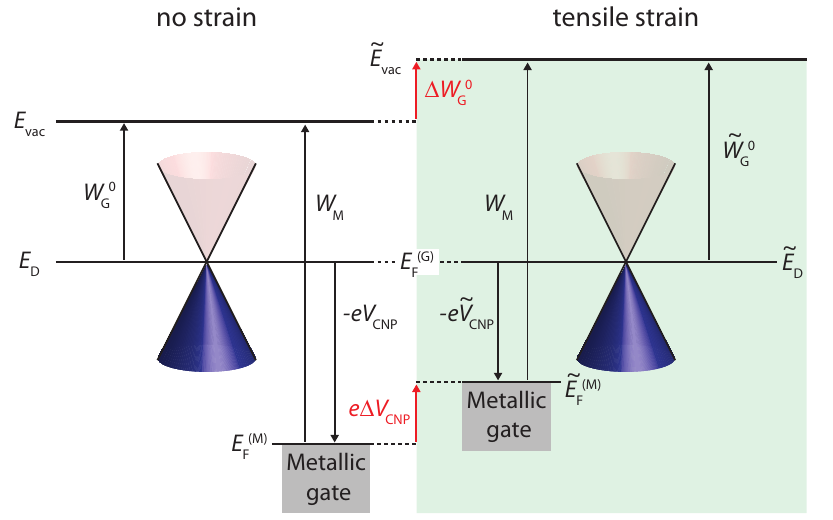}
	\caption{Schematic energy level diagram of the device at the CNP for unstrained (left) and strained (right, green shaded) graphene. The Fermi levels of the graphene and the metallic gate are denoted $E_{\rm F}^{(G)}$ and $E_{\rm F}^{(M)}$ ($\widetilde{E}_{\rm F}^{(M)}$), respectively. The WF of the metallic gate is denoted $W_{\rm M}$, assumed to be a constant. In our measurements, the graphene is grounded and therefore $E_{\rm F}^{(G)}$ is fixed. The gate voltages tuning the graphene to the CNP for the unstrained and strained cases are denoted $V_\text{CNP}$ and $\widetilde{V}_\text{CNP}$, respectively. The WF difference of undoped graphene between with and without strain is denoted $\mathrm{\Delta}W_{\rm G}^{0}$.}
	\label{fig:fig4}
\end{figure}

We now extract the scalar potential from the transport experiments by evaluating the shift between the minimum ($\mathrm{\Delta}z=0$) and maximum strain ($\mathrm{\Delta}z=\SI{0.8}{mm}$). We assume that a specific conductance feature, for example the CNP, or a QHE transition, occurs at a characteristic carrier density.
Here we use the CNP as an example for extracting the scalar potential. Figure~\ref{fig:fig4} shows the energy level alignment of the graphene gated to the CNP for the cases with and without strain. Different gate voltages are needed to gate the graphene to the CNP due to the strain-induced changes in the Dirac point energy (see Supplemental Material for details). At the CNP, the strain-induced scalar potential at $\mathrm{\Delta}z=\SI{0.8}{mm}$ can be directly extracted from Fig.~\ref{fig:fig4} and the inset of Fig.~\ref{fig:fig2}(b) as:

\begin{equation}\label{eq:S}
S^{\mathrm{\Delta}z=\SI{0.8}{mm}} = -\mathrm{\Delta}W_{\rm G}^{0} = -e\mathrm{\Delta}V_\text{CNP} \approx \SI{-8}{meV}.
\end{equation}

\begin{figure}[b]
	\centering
	\includegraphics[]{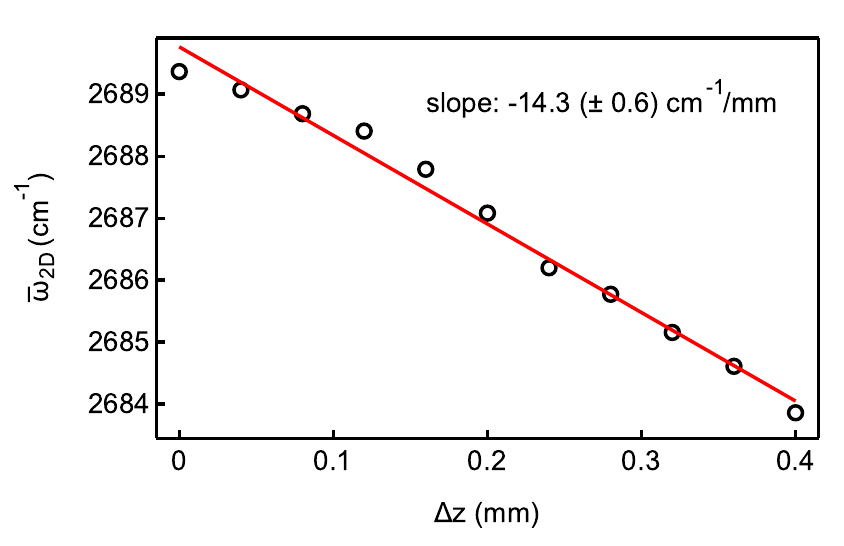}
	\caption{Spatially averaged center frequency $\bar{\omega}_\mathrm{2D}$ of the Raman 2D peak plotted as a function of $\mathrm{\Delta}z$. Black circles are data points and the red line is a linear fit. The linear decrease with increasing $\mathrm{\Delta}z$ indicates an increasing average strain.}
	\label{fig:fig5}
\end{figure}

To determine $s_0$ in Eq.~\ref{eq:Scalar_potential}, we need to estimate the applied strain. This we achieve using spatially resolved Raman spectroscopy at room temperature on the same device~\cite{Mohiuddin2009, Huang2010, Wang2019a}. For small uniaxial strain, a single Lorentzian describes the graphene Raman 2D peak, with the center frequency $\omega_\mathrm{2D}$ redshifting linearly with increasing tensile strain. Figure~\ref{fig:fig5} shows the mean center frequency $\bar{\omega}_\mathrm{2D}$ averaged over the entire device area as a function of $\mathrm{\Delta}z$. With increasing $\mathrm{\Delta}z$, $\bar{\omega}_\mathrm{2D}$ shifts to lower values, indicating an increasing average strain in the graphene~\cite{Wang2019a}. Since the displacement $\mathrm{\Delta}z$ is much smaller than the length of the substrate, the strain increases linearly with $\mathrm{\Delta}z$, with a slope of $\SI{\sim14.3}{\per\centi\meter}/\SI{}{\milli\meter}$ extracted by linear fitting. Using $\partial \omega_{2D}/\partial \varepsilon = \SI{-54}{\per\centi\meter}/\%$ from the literature~\cite{Mohr2009}, we obtain a value for the induced tensile strain of $\varepsilon = \varepsilon_{xx}+\varepsilon_{yy}\approx 0.21\%$ at $\mathrm{\Delta}z=\SI{0.8}{mm}$. With this calibration of the strain value, we now deduce the characteristic scalar potential constant $s_0 = - S/\varepsilon\approx \SI{3.8}{eV}$, which is within the range predicted by theory~\cite{Choi2010, Guinea2010, Grassano2020} and is consistent with the most recent calculations~\cite{Grassano2020}.

 

In conclusion, we have demonstrated how large scale homogeneous strain in a graphene electronic device results in a scalar potential, which we detect using transport experiments in two different regimes. Combined with strain values extracted from Raman spectroscopy on the same device, we report the first systematically measured characteristic number for the scalar potential strength, consistent with the most recent theoretical calculations. This \textit{in situ} strain tuning and the combination of transport and Raman measurements thus confirms the scalar potential as the origin of the observed strain effects. Our study forms the basis to investigate strain effects in transport experiments, which is crucial for future strain engineering in graphene and related 2D materials, such as generating a strain-induced in-plane electric field for observing the phenomenon of the Landau level collapse~\cite{Grassano2020}, realizing graphene quantum strain transistors~\cite{McRae2019}, or creating a pseudo-magnetic field with a non-uniform strain field~\cite{Guinea2009, Guinea2010}.

\subsection{Author contributions}
L.W. fabricated the devices, performed the measurements and did the data analysis. A.B., P.M., S.Z. and C.S. helped to understand the data. B.V. performed parts of the Raman measurements. D.I. supported the sample fabrication. K.W. and T.T. provided the high-quality hBN. C.S. initiated and supervised the project. L.W. and A.B. wrote the paper and all authors discussed the results and worked on the manuscript. All data in this publication are available in numerical form at: \href{https://doi.org/10.5281/zenodo.4017429}{https://doi.org/10.5281/zenodo.4017429}.

\section*{Acknowledgments}
This work has received funding from the Swiss Nanoscience Institute (SNI), the ERC project TopSupra (787414), the European Union Horizon 2020 research and innovation programme under grant agreement No. 785219 (Graphene Flagship), the Swiss National Science Foundation, the Swiss NCCR QSIT, Topograph, FlagERA network and from the OTKA FK-123894 grants. P.M. acknowledges support from the Bolyai Fellowship, the Marie Curie grant, Topograph Flagera network and the National Research, Development and Innovation Fund of Hungary within the Quantum Technology National Excellence Program (Project No. 2017-1.2.1-NKP-2017-00001). K.W. and T.T. acknowledge support from the Elemental Strategy Initiative conducted by the MEXT, Japan, Grant Number JPMXP0112101001, JSPS KAKENHI Grant Numbers JP20H00354 and the CREST(JPMJCR15F3), JST. The authors thank Francisco Guinea, Peter Rickhaus, J\'anos Koltai, L\'aszl\'o Oroszl\'any, Zolt\'an Tajkov and Andr\'as P\'alyi for fruitful discussions, and Sascha Martin and his team for their technical support.

\bibliography{ScalarPotential}

%


\end{document}


\begin{CJK*}{UTF8}{}

\title{Supplemental Material for \\Global strain-induced scalar potential in graphene devices}


\author{Lujun Wang}
\email{lujun.wang@unibas.ch}
\affiliation{Department of Physics, University of Basel, Klingelbergstrasse 82, CH-4056 Basel, Switzerland}
\affiliation{Swiss Nanoscience Institute, University of Basel, Klingelbergstrasse 82, CH-4056 Basel, Switzerland}

\author{Andreas Baumgartner}
\affiliation{Department of Physics, University of Basel, Klingelbergstrasse 82, CH-4056 Basel, Switzerland}
\affiliation{Swiss Nanoscience Institute, University of Basel, Klingelbergstrasse 82, CH-4056 Basel, Switzerland}

\author{P\'eter Makk}
\affiliation{Department of Physics, University of Basel, Klingelbergstrasse 82, CH-4056 Basel, Switzerland}
\affiliation{Department of Physics, Budapest University of Technology and Economics and Nanoelectronics Momentum Research Group of the Hungarian Academy of Sciences, Budafoki ut 8, 1111 Budapest, Hungary}

\author{Simon Zihlmann}
\affiliation{Department of Physics, University of Basel, Klingelbergstrasse 82, CH-4056 Basel, Switzerland}

\author{Blesson S. Varghese}
\affiliation{Department of Physics, University of Basel, Klingelbergstrasse 82, CH-4056 Basel, Switzerland}

\author{David I. Indolese}
\affiliation{Department of Physics, University of Basel, Klingelbergstrasse 82, CH-4056 Basel, Switzerland}

\author{Kenji Watanabe}
\affiliation{Research Center for Functional Materials, National Institute for Material Science, 1-1 Namiki, Tsukuba, 305-0044, Japan}

\author{Takashi Taniguchi}
\affiliation{International Center for Materials Nanoarchitectonics, National Institute for Materials Science, 1-1 Namiki, Tsukuba 305-0044, Japan}

\author{Christian Sch\"onenberger}
\affiliation{Department of Physics, University of Basel, Klingelbergstrasse 82, CH-4056 Basel, Switzerland}
\affiliation{Swiss Nanoscience Institute, University of Basel, Klingelbergstrasse 82, CH-4056 Basel, Switzerland}

\maketitle
\end{CJK*}

\section{Energy level alignment diagram}\label{sec:diagram}
\begin{figure}[H]
    \centering
      \includegraphics[width=\columnwidth]{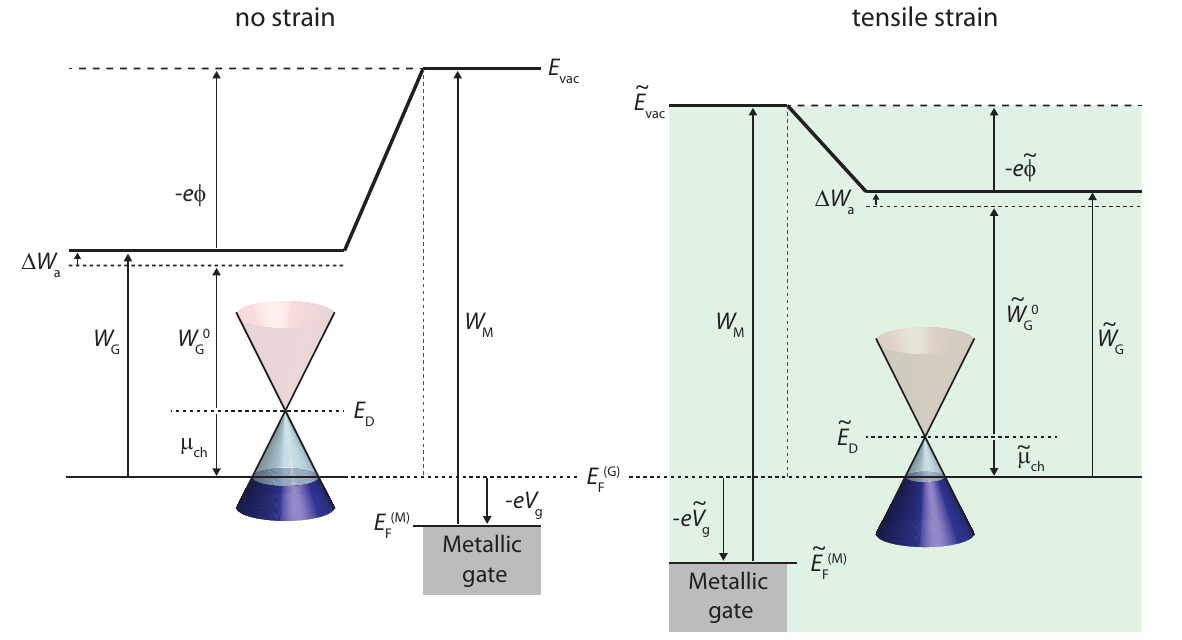}
    \caption{Schematic energy level diagram of the device for unstrained graphene at gate voltage $V_\text{g}$ (left) and for strained graphene at gate voltage $\widetilde{V}_\text{g}$ (right, green shaded). The symbols are defined in the text. In our measurements, the graphene is grounded through the metallic contacts and therefore $E_{\rm F}^{(G)}$ is fixed.}
    \label{fig:Diagram}
\end{figure}

The energy level diagram of a graphene electronic device at a gate voltage $V_{\rm g}$ is shown in Fig.~\ref{fig:Diagram}, where the band alignment between the metallic gate and the graphene connected to the grounded reservoirs is given by $E_{\rm F}^{(M)} - E_{\rm F}^{(G)} = - eV_{\rm g}$, with $E_{\rm F}^{(G)}$ and $E_{\rm F}^{(M)}$ the Fermi levels of the graphene and the metallic gate, respectively. Since the graphene is grounded via the metallic contacts, $E_{\rm F}^{(G)}$ is fixed throughout the measurements. 
The work function (WF) difference between the metallic gate ($W_{\rm M}$) and undoped graphene ($W_{\rm G}^{0}$) results in an electrostatic potential difference $\phi$ with a corresponding charge carrier density in the graphene. Due to the low density of states in graphene, one also needs to take into account the corresponding kinetic energies in the form of a finite chemical potential $\mu_{\rm ch}$, where $E_{\rm D}$ labels the Dirac point energy. Charged impurities in the surrounding, e.g. trapped charges in the dielectrics or adsorbed molecules nearby, induce an additional offset of the graphene WF, which we term $\Delta W_{\rm a}$~\cite{Giovannetti2008, Yu2009}. This offset then leads to a modification of $\mu_{\rm ch}$. In the end, the actual WF of the graphene ($W_{\rm G}$) in the device is $W_{\rm G} = \Delta W_{\rm a} + W_{\rm G}^{0} - \mu_{\rm ch}$. For the unstrained case, the fundamental relation between the relevant quantities can be directly found from Fig.~\ref{fig:Diagram}:
\begin{equation}\label{eq:W}
	\begin{split}
W_{\rm M} - eV_{\rm g} & = W_{\rm G} -e\phi \\ 
& = (\Delta W_{\rm a} + W_{\rm G}^{0} - \mu_{\rm ch}) -e\phi.
	\end{split}
\end{equation}
For tensile strain, the effect of the scalar potential is shown in the right part (shaded in green) of the diagram in Fig.~\ref{fig:Diagram}, where we label the resulting changed quantities by symbols with a tilde and assume that $W_{\rm M}$ is a constant (due to the very large density of states in metals) and $\Delta W_{\rm a}$ is not affected by strain. The strain-induced scalar potential shifts the Dirac point energy, leading to an increase in the intrinsic WF of undoped graphene, from $W_{\rm G}^{0}$ to $\widetilde{W}_{\rm G}^{0}$. This change has two effects in the device at a given gate voltage: a shift in the electrostatic potential difference, and a shift in the chemical potential. This results in a change in the charge carrier density, which can be detected in transport experiments. For the strained case at $\widetilde{V}_{\rm g}$, the corresponding quantities in Eq.~\ref{eq:W} should be replaced by those with a tilde, leading to 
\begin{equation}\label{eq:W_2}
	\begin{split}
W_{\rm M} - e\widetilde{V}_{\rm g} & = \widetilde{W}_{\rm G} -e\widetilde{\phi} \\ 
& = (\Delta W_{\rm a} + \widetilde{W}_{\rm G}^{0} - \widetilde{\mu}_{\rm ch}) -e\widetilde{\phi}.
	\end{split}
\end{equation}

To extract the changes in intrinsic WF of undoped graphene induced by strain, and therefore the strain-induced scalar potential, we observe the evolution of a given conductance feature, which we assume to occur at a specific charge carrier density, $n$.
A fixed $n$ corresponds to a fixed chemical potential ($\mu_{\rm ch} = \widetilde{\mu}_{\rm ch}$) in the graphene and a fixed electrostatic potential difference ($\phi = \widetilde{\phi}$), assuming no changes in the gate capacitance by straining~\cite{Wang2019a}. Under these conditions, the strain-induced scalar potential $S$ can be directly extracted from Eq.~\ref{eq:W} and ~\ref{eq:W_2}:
\begin{equation}\label{eq:S}
S = -(\widetilde{W}_{\rm G}^{0} - W_{\rm G}^{0}) = e(\widetilde{V}_{\rm g}^{f} - V_{\rm g}^{f}),
\end{equation}
where $V_{\rm g}^{f}$ and $\widetilde{V}_{\rm g}^{f}$ are the gate voltages for a given conductance feature for the unstrained and strained cases, respectively. This directly demonstrates that all conductance features are shifted by the same amount in gate voltage for a given global homogeneous straining, as observed in the experiments in the main text.

\clearpage
\section{Reversibility}\label{sec:reversibility}
\begin{figure}[H]
    \centering
      \includegraphics[width=\columnwidth]{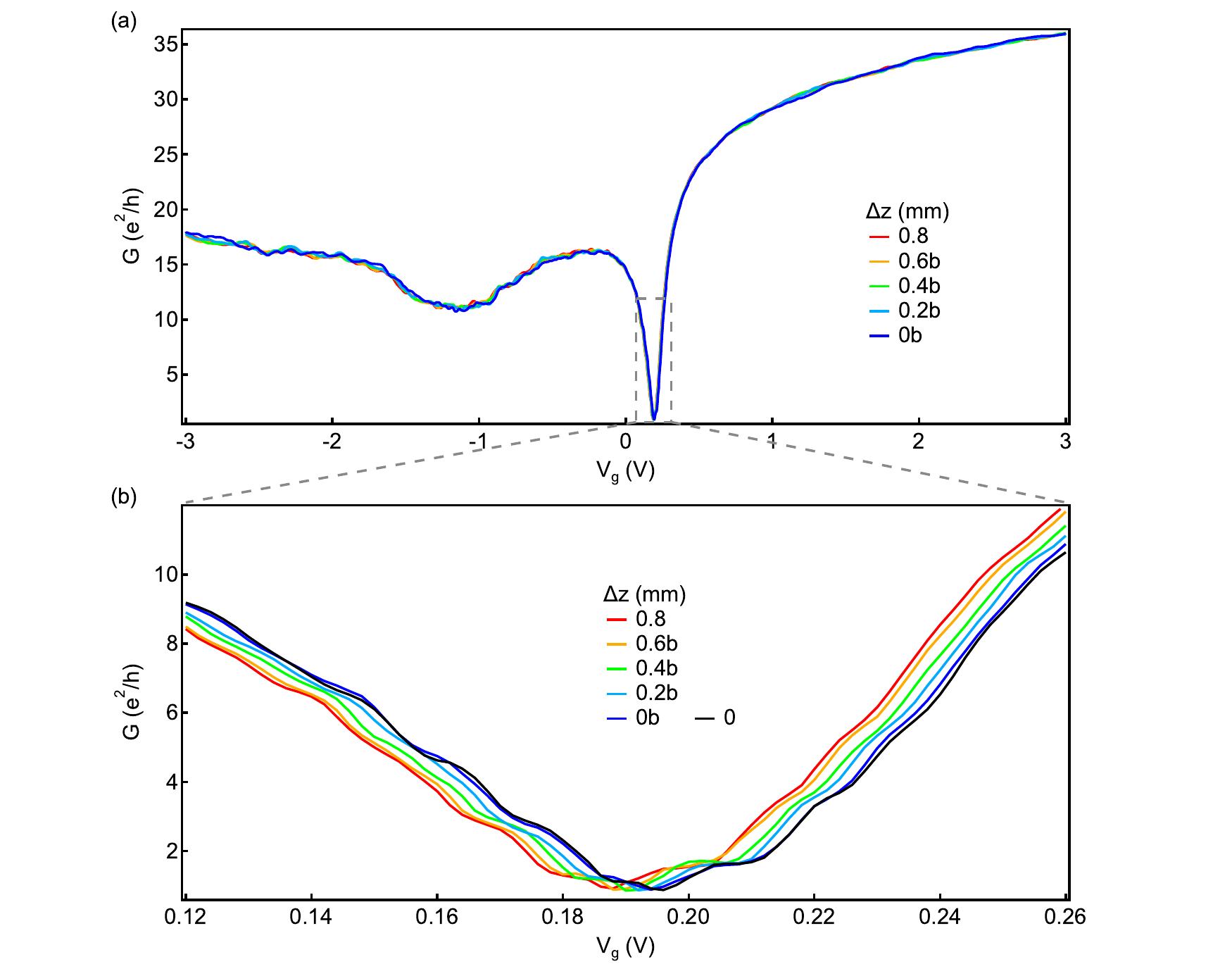}
    \caption{\textbf{(a)} Two-terminal differential conductance $G$ plotted as a function of gate voltage $V_\mathrm{g}$ for decreasing $\mathrm{\Delta}z$. ``b'' stands for ``backwards'' in $\mathrm{\Delta}z$. \textbf{(b)} Zoom-in to the CNP with the original $\mathrm{\Delta}z = 0$ curve added as black.}
    \label{fig:Supp_reversibility}
\end{figure}

Here we present the measurements of the device shown in the main text for decreasing $\mathrm{\Delta}z$ at zero magnetic field. The two-terminal differential conductance $G$ as a function of gate voltage $V_\mathrm{g}$ is shown in Fig.~\ref{fig:Supp_reversibility}(a) for different $\mathrm{\Delta}z$. No significant changes can be seen on this scale. An enlargement of the CNP is shown in Fig.~\ref{fig:Supp_reversibility}(b), where the original $\mathrm{\Delta}z = 0$ curve (black) is also added. With decreasing $\mathrm{\Delta}z$, the curve gradually reverts back to the original position, which demonstrates the full reversibility of the strain-induced scalar potential. 

\section{Discussion of the conductance oscillations}\label{sec:FP}
\begin{figure}[H]
    \centering
      \includegraphics[width=\columnwidth]{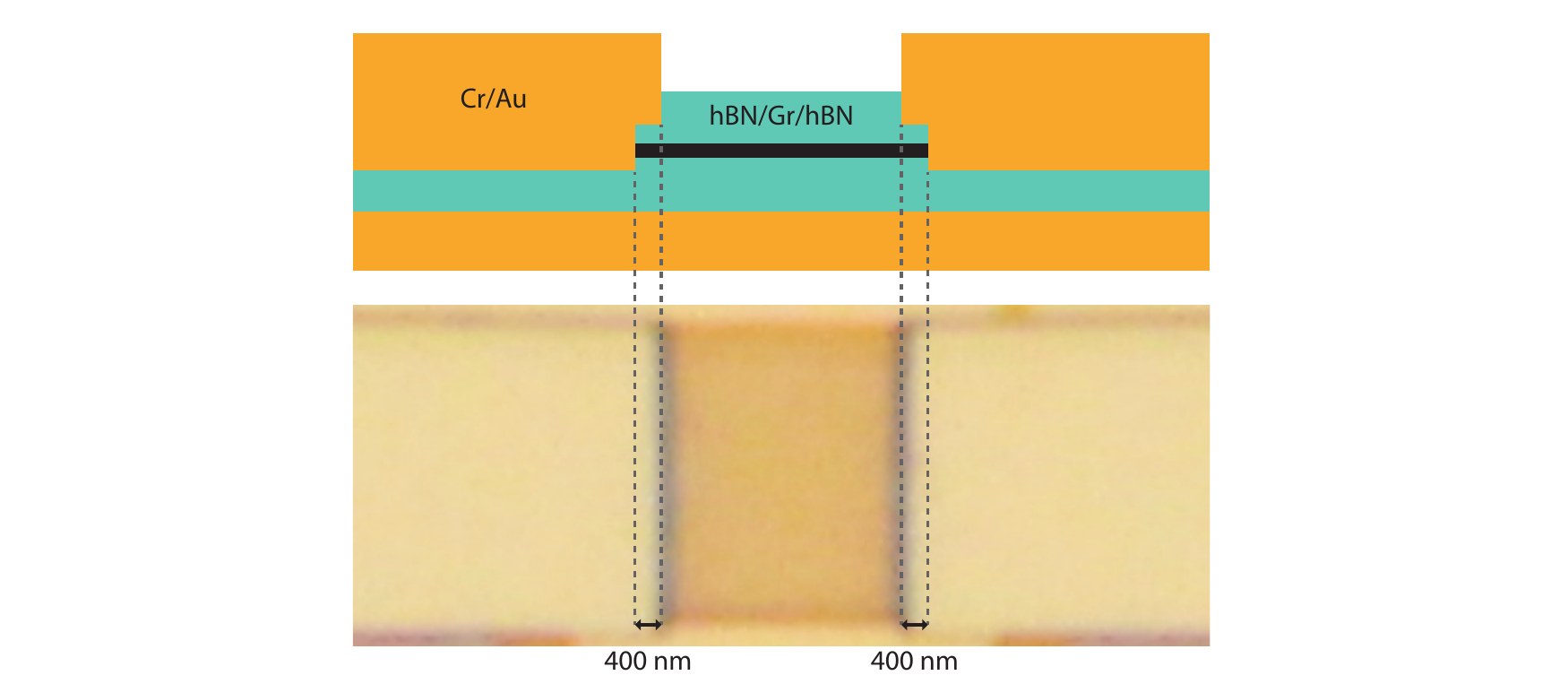}
    \caption{Schematic cross section and optical microscope image of the device. The \SI{400}{nm} overlap at each contact is intended for mechanical reinforcement.}
    \label{fig:Overlap}
\end{figure}

Very regular oscillations in conductance are observed around the CNP, which stay unaffected by strain, as shown in Fig.~\ref{fig:Supp_reversibility}(b). These features are reminiscent of the quantized conductance originating from quantum point contacts~\cite{Wees1988, Terres2016}. However, the corresponding Fermi wavelengths at the carrier densities of these oscillations are on the order of a few hundred \SI{}{nm}, estimated with $\lambda_\text{F}=2\pi/k_\text{F}=2\pi/\sqrt{\pi n}$ and $n = \alpha (V_\mathrm{g} - V_\mathrm{CNP})$ with $\alpha \approx \SI{4.9e15}{\per\square\meter\per\volt}$, which are an order of magnitude smaller than the device width (\SI{\sim4.4}{\micro\meter}). Therefore, the possibility of the quantized channel conductance as the origin of these oscillations is ruled out. Another possibility could be the Fabry-P\'erot resonances due to a region near each contact with a doping level different to the bulk of the device~\cite{Young2009, Rickhaus2013, Grushina2013, Handschin2017}. This is possible since the electrical contacts to the graphene are below the metallic leads and the overlap region is about \SI{400}{nm}, as shown in Fig.~\ref{fig:Overlap}. This design is intended for mechanical reinforcement of the contacts~\cite{Wang2019a}. From the conductance oscillations in the measurement, the cavity length $L$ can be estimated with $L=\sqrt{\pi}/(\sqrt{n_\text{j+1}}-\sqrt{n_\text{j}})$, where $\sqrt{n_\text{j+1}}$ and $\sqrt{n_\text{j}}$ are the corresponding carrier densities of two consecutive oscillations~\cite{Handschin2017}. A cavity length of \SI{\sim450}{nm} is extracted from our measurements, which matches well the \SI{\sim400}{nm} overlap near the contacts. We therefore tentatively attribute these oscillations to Fabry-P\'erot resonances. 

\section{Reversibility in quantum Hall regime}\label{sec:reversibility_QHE}
\begin{figure}[H]
    \centering
      \includegraphics[width=0.95\columnwidth]{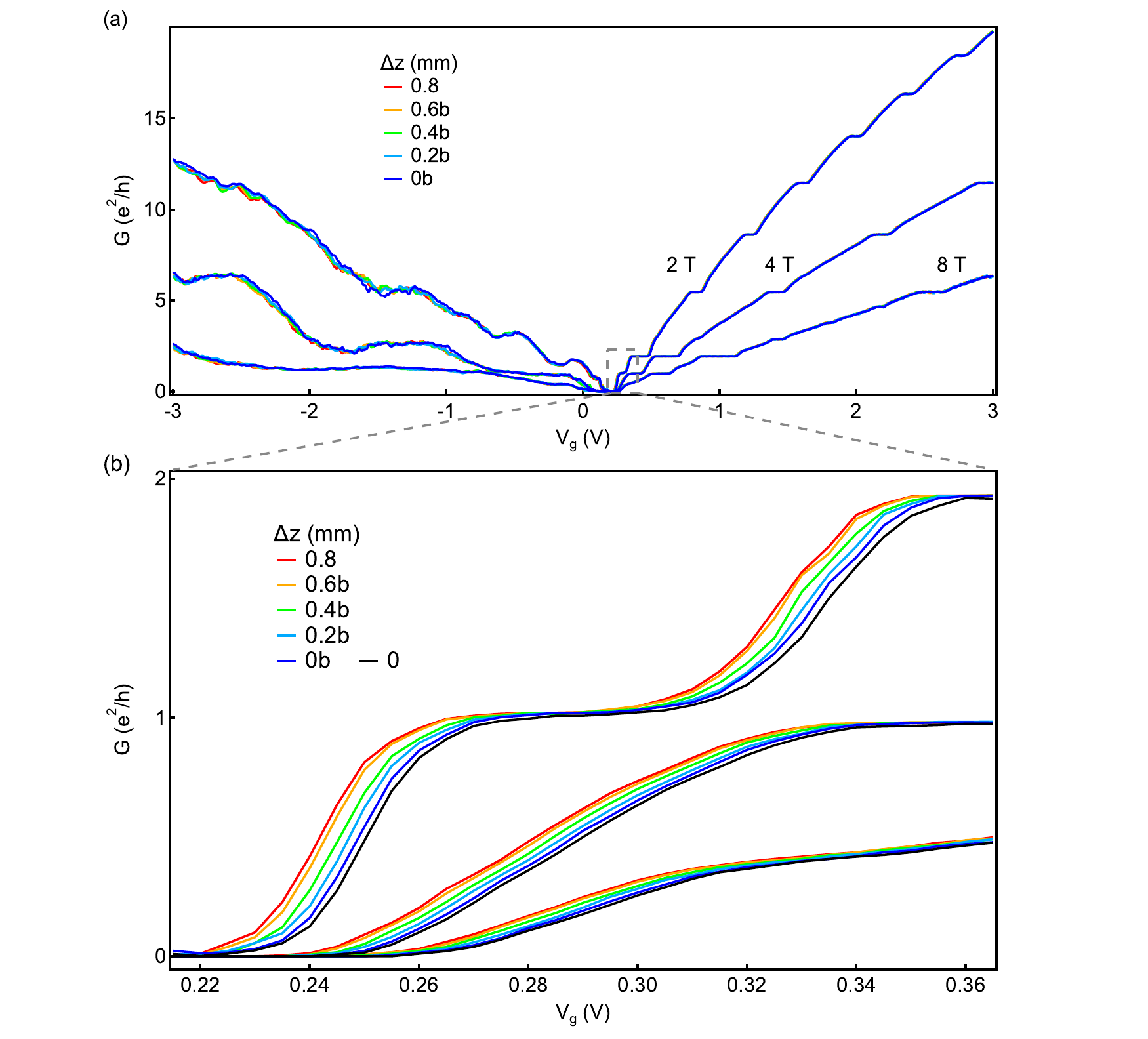}
    \caption{\textbf{(a)} Two-terminal differential conductance $G$ plotted as a function of the gate voltage $V_\mathrm{g}$ for decreasing $\mathrm{\Delta}z$ at $B=\SI{2}{T}$, $\SI{4}{T}$ and $\SI{8}{T}$. ``b'' stands for ``backwards'' in $\mathrm{\Delta}z$. \textbf{(b)} Zoom-in around the $\nu=1$ quantum Hall plateau with the original $\mathrm{\Delta}z = 0$ curve added as black.}
    \label{fig:Supp_reversibility_QHE}
\end{figure}

Figure~\ref{fig:Supp_reversibility_QHE}(a) shows the measured data from the same device as in the main text, but for decreasing strain values, at three different quantizing magnetic fields. No significant changes are found on this scale, except for the very left part. The quantum Hall plateaus in this region are not well developed due to a p-n junction forming near the contacts, which affects the electrical coupling of the graphene to the metallic contact. The changes observed here might be due to a small change in the effective contact resistance with strain. An enlargement around the $\nu = 1$ plateau is shown in Fig.~\ref{fig:Supp_reversibility_QHE}(b), where the shift of the curves in gate voltage due to strain-induced scalar potential can be seen. The curve gradually reverts back with decreasing $\mathrm{\Delta}z$, indicating the reversibility of the observed strain effect. The small mismatch between the $\mathrm{\Delta}z = 0$(b) curve (blue) and the original $\mathrm{\Delta}z = 0$ curve (black) can be attributed to a slight mechanical hysteresis of the bending setup.

\section{Second monolayer device showing scalar potential}
\label{sec:second}
\begin{figure}[H]
    \centering
      \includegraphics[width=0.95\columnwidth]{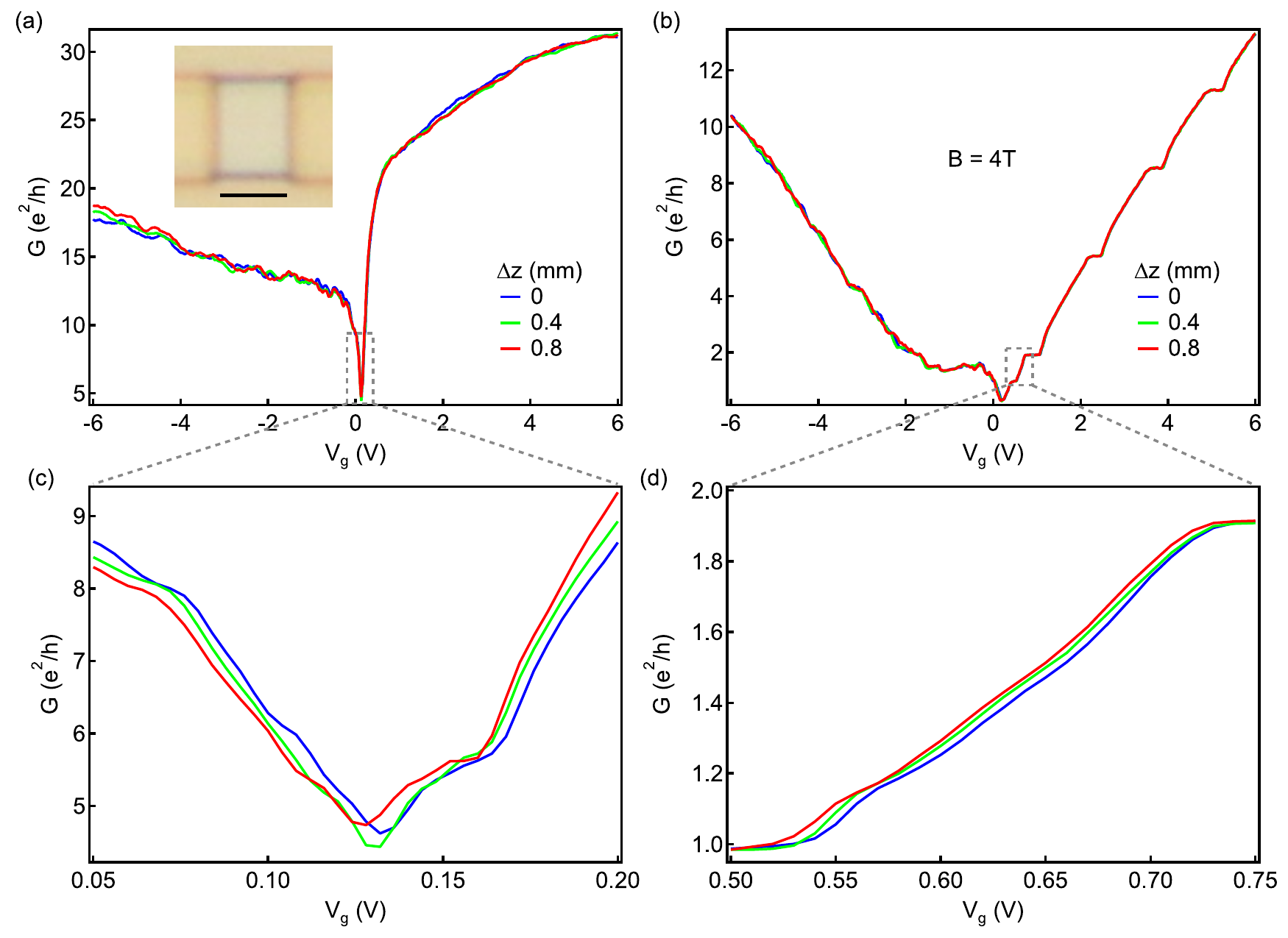}
    \caption{Two-terminal differential conductance $G$ plotted as a function of gate voltage $V_\mathrm{g}$ for different $\mathrm{\Delta}z$ at \textbf{(a)} $B = 0$ and \textbf{(b)} $B = \SI{4}{\tesla}$. The inset is an optical image of the measured device. Scale bar: \SI{2}{\micro\meter}. The corresponding enlargements near the CNP are shown in \textbf{(c)} and \textbf{(d)}, respectively.}
    \label{fig:CuBJ14J1}
\end{figure}
Data for a second monolayer device showing the strain-induced scalar potential are given here as another example. The conductance curves without magnetic field and in the quantum Hall regime are plotted in Fig.~\ref{fig:CuBJ14J1}(a) and (b), respectively. No significant effects are observed with increasing $\mathrm{\Delta}z$ on this scale, except for the very left part of Fig.~\ref{fig:CuBJ14J1}(a), where the conductance is limited by the p-n junction near the contact. The changes seen in this region might be attributed to a small change in the contact resistance, which does not affect the interpretation of the strain effect observed near the CNP, as shown in Fig.~\ref{fig:CuBJ14J1}(c) and (d). A shift of $\SI{\sim7}{mV}$ to the left in $V_\mathrm{g}$ is observed for $\mathrm{\Delta}z = \SI{0.8}{mm}$, which is similar to that of the device shown in the main text, and is thus attributed to the strain-induced scalar potential.

\section{Fabrication and Raman measurements}\label{sec:Fab}
The hBN/graphene/hBN heterostructures were first assembled using the standard pick-up technique with a PDMS/PC stamp and then deposited onto the metallic gate structure prefabricated on a polyimide-coated phosphor bronze plate. The typical thickness for the top (bottom) hBN is \SI{\sim20}{\nano\meter} (\SI{\sim30}{\nano\meter}). The graphene flake was exfoliated from natural graphite. One-dimensional edge contacts~\cite{Wang2013} (Cr/Au, \SI{5}{\nano\meter}/\SI{110}{\nano\meter}) were made to electrically connect the graphene. A controlled etching recipe was employed to stop in the middle of the bottom hBN and the remaining hBN acts as the insulating layer between the metallic leads and the bottom gate~\cite{Wang2019a}, as can be seen in Fig.~\ref{fig:Overlap}.

The Raman measurements at room temperature were performed to determine the strain after the low-temperature transport measurements. A commercially available confocal Raman system WiTec alpha300 was used. The Raman spectra were acquired using a linearly polarized green laser ($\SI{532}{\nano\meter}$) with a power of $\SI{1.5}{\milli\watt}$. The grating of the spectrometer is 600 grooves/mm.

\bibliography{ScalarPotential}